\documentclass[10pt,
nofootinbib,showpacs,reprint,
aps,prd]{revtex4-1}
\usepackage{amsfonts,amsmath,amssymb,mathrsfs}
\usepackage{mathtools,mathptmx,array,url}
\usepackage{appendix}
\usepackage{subfigure,graphics,graphicx}
\usepackage[colorlinks=true,linkcolor=blue,citecolor=blue,urlcolor=blue]{hyperref}
\usepackage{ulem}
\def\be{\begin{equation}}
\def\ee{\end{equation}}
\def\bea{\begin{eqnarray}}
\def\eea{\end{eqnarray}}
\def\beal{\begin{aligned}}
\def\eeal{\end{aligned}}
\def\nn{\nonumber}
\def\p{\partial}
\def\cA{\mathcal{A}}

\begin{document}

\title{
Shadowless rapidly rotating yet not ultraspinning Kerr-AdS$_4$ and Kerr-Newman-AdS$_4$ black holes}

\author{Di Wu$^{1,2}$}
\email{wdcwnu@163.com}

\author{Puxun Wu$^{2}$}
\email{ pxwu@hunnu.edu.cn}

\author{Hongwei Yu$^{2}$}
\email{ hwyu@hunnu.edu.cn}

\affiliation{
$^{1}$College of Physics and Space Science, China West Normal University, Nanchong,
Sichuan 637002, People's Republic of China \\
$^{2}$Department of Physics and Synergetic Innovation Center for
Quantum Effect and Applications, Hunan Normal University, Changsha, Hunan 410081,
People's Republic of China 
 }

\date{\today}

\begin{abstract}
We find that the Kerr-(Newman)-AdS$_4$ black hole will be shadowless  if its rotation parameter is larger than a critical value   and the shadowless-ness  may be related to the appearance of the null hypersurface caustics (NHC)   both inside the Cauchy horizon and outside the event horizon for the black hole with the rotation parameter beyond the critical value.    Our studies also further confirm that whether an ultraspinning black hole is super-entropic or not is unrelated to the existence of the NHC outside the event horizon.

\end{abstract}

\maketitle

The black hole, as one of the most remarkable and fascinating objects in nature,  has always been a subject of extensive studies both theoretically and observationally. 
On the observational side, recent years have witnessed exciting achievements in ``hearing" the gravitational waves from the coalescence of  black holes~\cite{PRL116-061102,PRL116-241103,PRX9-031040,PRL125-101102,PRX11-021053,APJ915-L5} as well as ``seeing"
its shadow by  the event horizon telescope~\cite{APJ875-L1,APJ875-L2,APJ875-L3,APJ875-L4,
APJ875-L5,APJ875-L6,APJ910-L12,APJ910-L13}. 
 On the theoretical side, the study of  asymptotically anti de-Sitter (AdS) black holes has shed light on the nature of gravity through gauge-gravity dualities \cite{ATMP2-231,PLB428-105,ATMP2-253}. About ten years ago, it was conjectured that an AdS black hole should  satisfy the reverse isoperimetric inequality (RII) \cite{PRD84-024037,PRD87-104017}
\bea\label{RII}
\mathcal{R} = \Big[\frac{(d-1)V}{\cA_{d-2}}\Big]^{1/(d-1)}\Big(\frac{\cA_{d-2}}{A}\Big)^{1/(d-2)}
\ge 1 \, ,
\eea
where $V$ is the thermodynamic volume of the black hole, $\cA_{d-2} = 2\pi^{[(d-1)/2]}/\Gamma[(d-1)/2]$
 the area of the unit $(d-2)$-sphere with $d$ being the number of spacetime dimensions and $A$ 
the area of the outer horizon. Equality is attained for the Schwarzschild-AdS black hole, which
means that the Schwarzschild-AdS black hole has the maximum entropy or it occupies the least volume for a specified entropy.

The first counterexample of an AdS black hole, which does not satisfy the RII, is the ultraspinning  black hole~\cite{PRL115-031101},  which is constructed by boosting one of the angular velocities of the rotating AdS black hole to the speed of light~\cite{PRL115-031101}.  Such constructed ultraspinning black hole has a finite horizon area but noncompact event horizon topology. Since the ultraspinning   black hole exceeds the maximum entropy bound~\cite{PRD89-084007,JHEP0114127,JHEP0220195}, it is dubbed ``superentropic". Soon afterward, a lot of new ultraspinning AdS black hole solutions \cite{JHEP0615096,PRD95-046002,1702.03448,JHEP0118042,PRD102-044007,PRD103-044014,2106.14218} have been generated successfully from the corresponding  known rotating AdS black holes. And various aspects of the ultraspinning black holes, including thermodynamic properties \cite{PRL115-031101,JHEP0615096,PRD95-046002,1702.03448,JHEP0118042,2106.14218,MPLA35-2050098,PRD101-086006,
PRD101-024057,PLB807-135529}, horizon geometry \cite{PRD89-084007,JHEP0615096,PRD95-046002},
Kerr/CFT correspondence \cite{PRD95-046002,1702.03448,JHEP0816148} and geodesic motion \cite{JMP61-122504}, etc, have been explored widely.  However, not all ultraspnning black holes are superentropic since  a few new ultraspinning AdS black holes \cite{JHEP0615096, PRD102-044007,PRD103-044014}, i.e. the ultraspinning Kerr-Sen-AdS$_4$ (dyonic) black holes, violate the RII only in certain parameter spaces of the solutions. Therefore, ultraspinning is not a sufficient condition for a black hole to be superentropic. 

So, a question arises naturally as to what the physical condition is for an ultraspinning black hole to be superentropic. Recently,  it was suggested that the superentropic-ness may be related to the presence of the null hypersurface caustics (NHC) outside the event horizon of the black hole. This is because  the NHC exists outside the event horizon for  the (high dimensional) Kerr-(Newman)-AdS superentropic black hole~\cite{CQG38-045018,PRD103-104020}. The existence of the NHC  is thought previously to mean that the causal structure of spacetime has some pathologies. However, it was recently found that the spacetime with the NHC is free of closed timelike curve \cite{PRD103-024053}.  

For the ultraspinning Kerr-Sen-AdS$_4$ black hole, whether it is superentropic or not, the NHC always appears both out and inside of the horizon \cite{PRD103-024053}, which seems to indicate that the existence of the NHC outside the event horizon is related to ultraspinning rather than  superentropicness. The ultraspinning means that the rotational angular velocity of the black hole is boosted to the speed of  light.  Then a further issue, which we are going to address, is whether the rotational  angular velocity reaching the speed of light is a necessary requirement for existence of the NHC  outside the event horizon.

\begin{figure}[!t]
\centering
\includegraphics[width=0.4\textwidth]{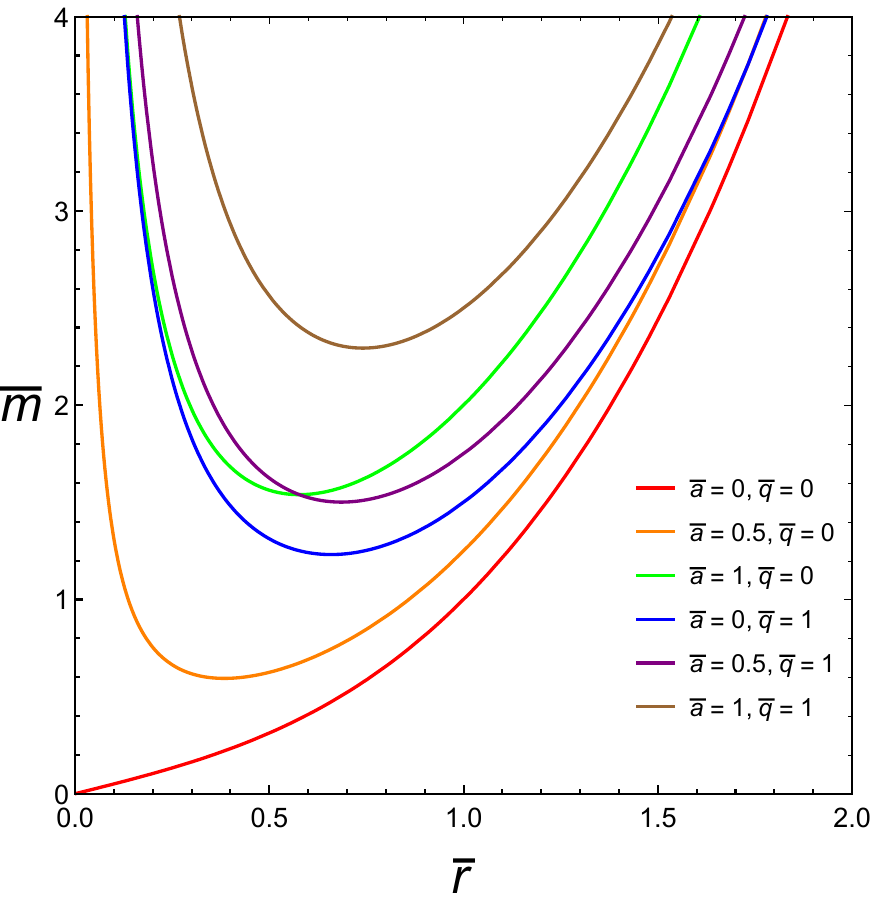}
\caption{The rescaled mass vs the rescaled horizon radius, where $\bar{m} = m/l$, $\bar{a} = a/l$, $\bar{q} = q/l$, $\bar{r} = r/l$. The red, orange, green, blue, purple and brown lines represent the Schwarzschild-AdS$_4$, Kerr-AdS$_4$, ultraspinnning Kerr-AdS$_4$, RN-AdS$_4$, Kerr-Newman-AdS$_4$ and ultraspinning Kerr-Newman-AdS$_4$ black holes, respectively.
\label{mlb}}
\end{figure}

Let us start with the four-dimensional Kerr-Newman-AdS black hole solution \cite{CMP10-280}, whose metric has the form~\cite{PRD59-064005}
\bea\label{KNAdS}
ds^2 &=& -\frac{\Delta_r}{\Sigma}\left(dt -\frac{a}{\Xi}\sin^2\theta d\phi \right)^2 +\frac{\Sigma}{\Delta_r}dr^2 +\frac{\Sigma}{\Delta_\theta}d\theta^2 \nn \\
&&+\frac{\Delta_\theta\sin^2\theta}{\Sigma}\left(adt -\frac{r^2 +a^2}{\Xi}d\phi \right)^2 \,
\eea
in the Boyer-Lindquist coordinates, where
\bea
&&\Delta_r = (r^2 +a^2)\left(1 +r^2/l^2\right) -2mr +q^2 \, , \quad \Xi = 1 -a^2/l^2 \, , \nn \\
&&\Delta_\theta = 1 -a^2\cos^2\theta/l^2 \, , \quad \Sigma = r^2 +a^2\cos^2\theta \, , \nn
\eea
$a$ is the rotation parameter, $m$  the mass parameter, $q$  the electric charge parameter and $l$  the AdS radius. The horizon is determined by equation  $\Delta_r = 0$. When $q = 0$,  the  metric  in Eq.~(\ref{KNAdS}) reduces to that of the Kerr-AdS$_4$ black hole. The RN-AdS$_4$ black hole is obtained if  $a=0$ in Eq.~(\ref{KNAdS}).  In Fig. \ref{mlb}, we plot the rescaled horizon radius $\bar{r}$ as a function of the scaled mass $\bar{m}$. It is easy to see that if the black hole mass is larger than a critical value both the RN-AdS$_4$ black hole and the  Kerr-(Newman)-AdS$_4$ black hole have the Cauchy and event horizons. When  the black hole mass equals to  the critical value, the Cauchy and event horizons coincide.  While the Schwarzschild-AdS$_4$ black hole, which corresponds to the case with both $a=0$ and $q=0$,  only has the event horizon.

To derive the condition for existence of  the NHC for the Kerr-Newman-AdS$_4$ black hole, we  use the method given in  Refs. \cite{CQG15-2289,CQG36-245017,CQG38-045018,PRD103-024053,PRD103-104020}. After  introducing  the outgoing and ingoing Eddington-Finkelstein coordinates defined in terms of the ``generalized tortoise coordinate" $r_*(r,\theta)$: $u = t - r_*(r,\theta)$, $v = t +r_*(r,\theta)$, the null hypersurfaces can be described by $u = {\rm const}$, $v = {\rm const}$, which are dubbed the outgoing and ingoing null congruences of the hypersurfaces, respectively. It is easy to obtain that  the null hypersurfaces defined by $u = {\rm const}$ or $v = {\rm const}$ satisfy the equation
\be\label{PDE}
g^{\mu\nu}\p_\mu(t +r_*)\p_\nu(t +r_*) = g^{tt} +g^{rr}(\p_r r_*)^2 +g^{\theta\theta}(\p_\theta r_*)^2 = 0 \, .
\ee
By using the contravariant components $g^{tt}$,  $g^{rr}$, and $g^{\theta\theta}$ of the metric (\ref{KNAdS}), Eq.~(\ref{PDE}) can be re-expressed as
\be\label{PDE2}
\Delta_r\left(\p_r r_* \right)^2 - \Xi^2\frac{(r^2 +a^2)^2}{\Delta_r}   = -\Delta_\theta\left(\p_\theta r_* \right)^2- \Xi^2 \frac{a^2\sin^2\theta}{\Delta_\theta}   \, . \quad
\ee
This equation can be solved by  the method of separation of variables. Introducing the so-called ``constant of separation" $a^2\lambda$ (hereinafter, $\lambda$ is referred to as the separation constant) for Eq.~(\ref{PDE2}),  we have
\be\label{rr2}
(\p_r r_*)^2 = \frac{Q^2(r)}{\Delta_r^2} \, , \quad (\p_\theta r_*)^2 = \frac{P^2(\theta)}{\Delta_\theta^2} \, .
\ee
where
\be\beal\label{Q2P2}
&Q^2(r) = \Xi^2\left[\left(r^2 +a^2 \right)^2 -a^2\lambda\Delta_r \right] \, ,  \\
&P^2(\theta) = \Xi^2a^2\left(\lambda\Delta_\theta -\sin^2\theta \right) \, .
\eeal
\ee
 In \cite{CQG38-045018,CQG36-245017},  $\lambda$  is chosen as  $\lambda<1$ and it was then found that  the NHC cannot  exist outside the event horizon of the Kerr(-Newman)-AdS black hole. However, this choice  is incorrect, since when $\theta=\pi/2$, $P^2(\theta) = \Xi^2a^2\left(\lambda -1 \right) )\geq 0$ must be satisfied. Thus,  the allowed region of $\lambda$ should be  $\lambda \ge 1$.

\begin{figure}[!t]
\centering
\includegraphics[width=0.4\textwidth]{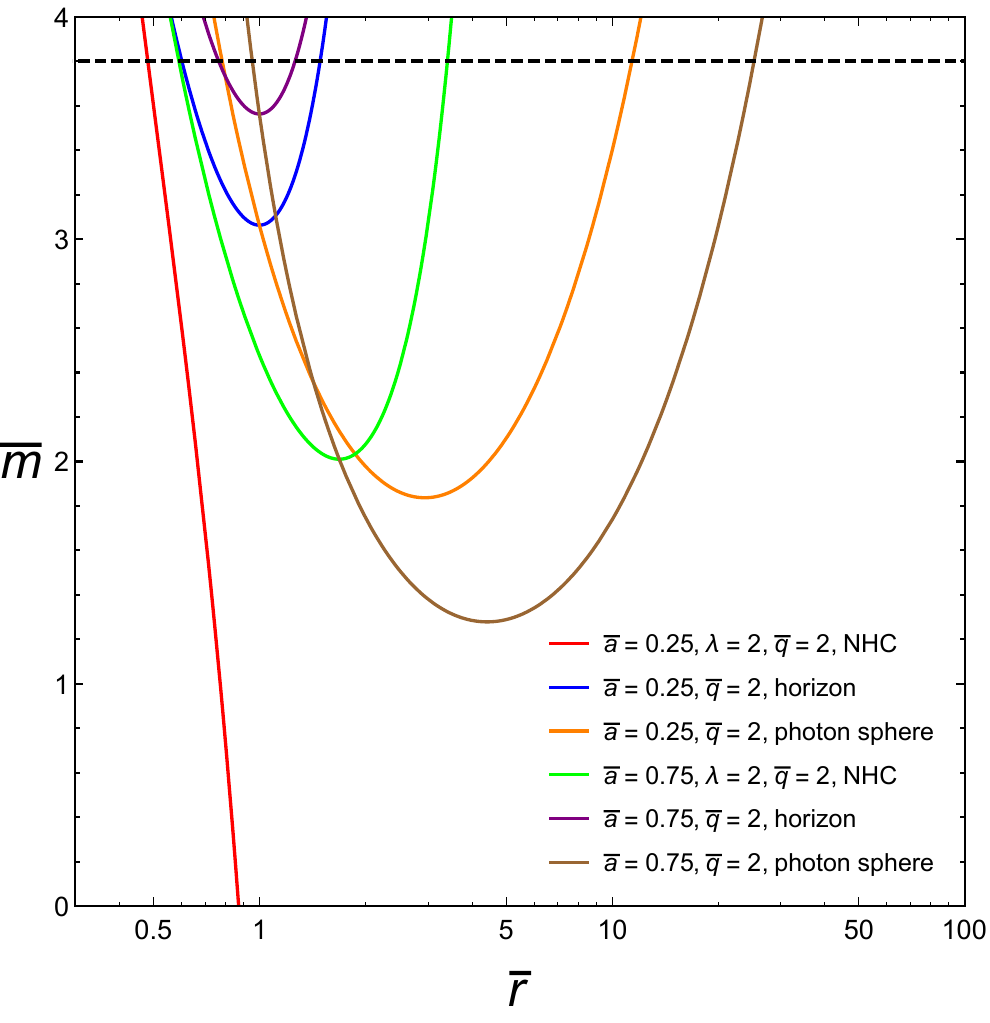}
\caption{Contours of the NHC, the horizon and the photon sphere in the ($\bar{r}$, $\bar{m}$) plane with a fixed $\bar{q}$ for the Kerr-Newman-AdS$_4$ black hole, where $\bar{m} = m/l$, $\bar{a} = a/l$, $\bar{q} = q/l$, $\bar{r} = r/l$.
\label{NHCps}}
\end{figure}

After some complicated computations, which are presented in detail in the Appendix, we find that $Q(r) = 0$ is a sufficient condition for the existence of the NHC, which means that
\be\label{ccKNAdS}
\left(r^2 +a^2 \right)^2 -a^2\lambda\Delta_r = 0 \, .
\ee
Figure \ref{NHCps} shows the numerical results of the NHC condition (\ref{ccKNAdS}).  In addition,  the photon sphere condition, which is governed by the equation $(r^2 +a^2)\Delta_r^{\prime} -4r\Delta_r = 0$~\cite{PRD89-124004,2007.09058}, and the horizon condition ($\Delta_r = 0$), are plotted, where a prime indicates the derivative  with respect to $r$.  In this figure, two intersection points between the dashed black line and the blue/purple line represent that the black hole has  the Cauchy horizon and the event horizon. One can see that the NHC exists only inside the Cauchy horizon when $\bar{a} = 0.25$, and it appears both inside the Cauchy horizon and outside the event horizon of the black hole when $\bar{a} = 0.75$.  This suggests that the NHC can appear outside the event horizon of the black hole for a  large enough  $a$ which is not necessarily  equal to $l$. The photon sphere line indicates that the NHC, which appears outside of the event horizon, locates between the event horizon and the photon sphere of the black hole.

How large is the value of the rotational angular velocity  required for existence of the NHC outside the event horizon? After a detailed calculation, we find that when the rotation parameter $a$ satisfies
\be\label{ine}
a \ge \frac{l}{\sqrt{\lambda}} \, ,
\ee
the NHC exists both outside the event horizon and inside the Cauchy horizon of the Kerr-(Newman)-AdS$_4$ black hole.  Otherwise, there is the NHC only inside the Cauchy horizon. Thus, if the rotation parameter is larger than a critical value $a_c= {l}/{\sqrt{\lambda}}$, the NHC appears outside the event horizon. Apparently, this critical value is determined by $l$ and $\lambda$. When $\lambda$ equals unity, the NHC exists outside the event horizon  only in the case of the ultraspinning Kerr-(Newman)-AdS$_4$ black hole ($a \to l$). If $\lambda>1$, a large enough $a$, which is not necessarily equal to $l$, can lead to the existence of the NHC outside the event horizon of the black hole.  Therefore,  the value of $\lambda$ determines how large $a$ has to be to ensure the existence of the NHC outside the event horizon.


%

\begin{figure}[!h]
\centering
\includegraphics[width=0.45\textwidth]{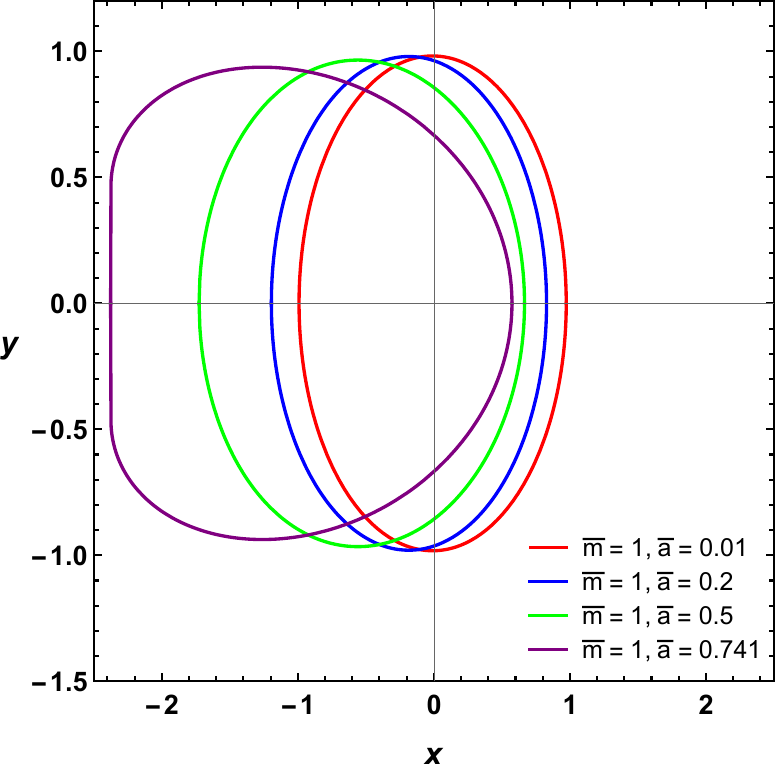}
\caption{The shadows for the Kerr-AdS$_4$ black hole for different rescaled rotation parameter $\bar{a}$ with $\bar{m}=1$, where $\bar{m} = m/l$, $\bar{a} = a/l$.
\label{shadow}}
\end{figure}

    Now we demonstrate that when the rotation parameter is larger than a critical value,  in addition to the existence of the NHC outside the event horizon,
   the Kerr-(Newman)-AdS black hole's shadow  cannot be obtained  and thus it should be shadowless.  Here, we take the Kerr-AdS$_4$ black hole as an example and investigate its shadows.  The results are expected to be valid for the Kerr-Newman-AdS$_4$ case.   Figure \ref{shadow} gives the numerical results in the equatorial plane $\theta = \pi/2$. With the increase of the rotational angular velocity, the black hole shadows change from an ellipse to a ``D" shape. Once $a > 0.741l$, the shadow of the Kerr-AdS$_4$ black hole disappears. 
    Clearly,  both the shadowless-ness and the existence of the NHC outside the event horizon of the Kerr-(Newman)-AdS black hole require that the black hole rotates rapidly enough with the rotation parameter larger than a critical value.  This seems to suggest that the existence of the HNC outside the event horizon might be the cause  of the shadowless-ness of a rapidly rotating Kerr-(Newman)-AdS black hole.  Therefore, we can obtain the value of $\lambda$ from analyzing the shadow of the Kerr-(Newman)-AdS$_4$ black hole. From the inequality (\ref{ine}), we find that the value of $\lambda$ should be chosen to be about $\lambda \sim 1.821$ when $\bar{m}=1$.

It is easy to obtain that when the rotation parameter of the black hole is larger than the critical value $a_{c}$, i.e. $a_c\simeq 0.741 l$ for the Kerr-AdS$_4$ black hole with $\bar{m}=1$,  the NHC will appear outside the event horizon of the black hole. Apparently, the ultraspinning limit  ($a\rightarrow l$) of the black hole is not a necessary but sufficient  condition for the appearance of the NHC outside the event horizon of black hole. Thus,  we further confirm that the superentropic-ness of the black hole  is  unrelated to the presence of the NHC outside the  event horizon of the rotating AdS black hole.


To summarize, by investigating the NHC of the Kerr-(Newman)-AdS$_4$ black hole, we find that there exists a critical value $a_{c}$ for the rotation parameter $a$. When the rotation parameter of the black hole is smaller than this critical value, the NHC only exists inside the Cauchy horizon.  Once  $a$ is larger than this critical value,  the Kerr-(Newman)-AdS$_4$ black hole will have the NHC both inside the Cauchy horizon and outside the event horizon.  Our results demonstrate that the NHC can exist outside the event horizon of  the black hole which is not ultraspinning,  in sharp contrast to the conclusion that the Kerr-(Newman)-AdS$_4$ black hole has  the NHC outside the event horizon only when it is ultraspinning~\cite{CQG38-045018,CQG36-245017}.     The  fact that the shadows of the rotating AdS black holes cannot be obtained  if the rotation parameter is larger than a critical value seems to suggest that the appearance of the NHC outside the event horizon leads to the disappearance of the black hole shadow. Thus, for the rapidly rotating yet not ultraspinning  the Kerr-(Newman)-AdS$_4$ black holes,  we will not see their shadows neither.  Our results   confirm that the superentropic-ness of ultraspinning black holes is unrelated to the presence of the NHC outside the event horizon. Finally, we believe that  our work  can be generalized to other rotating AdS black holes, and for these black holes there is also a critical value for the rotation parameter.

\acknowledgments
We thank Professor Shuang-Qing Wu and Mr. Yuebing Zhou for helpful discussions. This work is supported by the National Natural Science Foundation of China under Grants No. 11675130, No. 11775077, No. 11690034, No. 12075084, by the National Key Research and Development Program of China Grant No. 2020YFC2201502, and by the Science and Technology Innovation Plan of Hunan province under Grant No. 2017XK2019.


\appendix
\begin{appendices}
\section{} 
Using the relations given in (5), one can express the total differential $dr_* = \p_r r_*dr +\p_\theta r_*d\theta$ as
\be\label{dr}
dr_* = \frac{Q(r)}{\Delta_r}dr +\frac{P(\theta)}{\Delta_\theta}d\theta \, .
\ee
By treating $\lambda$ as a variable, the exact differential Eq.~(\ref{dr}) can be generalized to be
\be\label{dr1}
dr_* = \frac{Q(r,\lambda)}{\Delta_r}dr +\frac{P(\theta,\lambda)}{\Delta_\theta}d\theta +c_1F(r,\theta,\lambda)d\lambda \, ,
\ee
where $c_1$ is an arbitrary constant and $F(r,\theta,\lambda)$ is an arbitrary function. Since Eqs.~(\ref{dr}) and (\ref{dr1}) are functionally equivalent, the condition
\be
F(r,\theta,\lambda) = 0
\ee
should be  satisfied, which indicates $dF(r,\theta,\lambda) = 0$. Thus, we arrive at 
\be\label{dF}
[\p_\lambda F(r,\theta,\lambda)]d\lambda +[\p_r F(r,\theta,\lambda)]dr
 +[\p_\theta F(r,\theta,\lambda)]d\theta
= 0 \, .
\ee
From the Poincar\'e lemma in the external differentiation theory, namely $d(dr_*) = 0$, one can
obtain the following integrable conditions
\be
\frac{\p_\lambda Q(r,\lambda)}{\Delta_r} = c_1\p_rF(r,\theta,\lambda)\, , \quad
\frac{\p_\lambda P(\theta,\lambda)}{\Delta_\theta} = c_1\p_\theta F(r,\theta,\lambda) \, .
\ee
From the definitions of $Q(r, \lambda)$ and $P(\theta, \lambda)$ given in (6), one finds straightforwardly
\be
\p_\lambda Q(r,\lambda) = -\frac{\Xi^2a^2\Delta_r}{2Q(r,\lambda)} \, , \quad
\p_\lambda P(\theta,\lambda) = \frac{\Xi^2a^2\Delta_\theta}{2P(\theta,\lambda)} \, ,
\ee
and one can rewrite Eq. (\ref{dF}) as
\be\label{nu}
\nu d\lambda = -\frac{dr}{Q(r,\lambda)} +\frac{d\theta}{P(\theta,\lambda)} \, ,
\ee
after choosing the constant $c_1$ to be $\Xi^2a^2/2$ and defining $\nu = -\p_\lambda F(r,\theta,\lambda)$. By making use of Eq. (\ref{dr}) and Eq. (\ref{nu}), the metric of the Kerr-Newman-AdS$_4$ metric given in (2) can be re-expressed in terms of the coordinates ($t,r_*,\theta,\lambda$)
\bea
ds^2 &=& \frac{\Delta_r\Delta_\theta}{\Xi^4R^2}(dr_*^2 -dt^2) +R^2\sin^2\theta(d\phi -\Omega dt)^2  \nonumber \\ &&\quad+\frac{\nu^2P^2(\theta,\lambda)Q^2(r,\lambda)}{\Xi^4R^2}d\lambda^2 \, ,
\eea
where
\bea
&&R^2 = \frac{g_{\phi\phi}}{\sin^2\theta} = \frac{\Delta_\theta(r^2 +a^2)^2 -\Delta_ra^2\sin^2\theta}{\Xi^2\Sigma} \, , \\
&&\Omega = -\frac{g_{t\phi}}{g_{\phi\phi}} = \frac{a\left[\Delta_\theta(r^2 +a^2) -\Delta_r \right]}{\Xi\Sigma R^2} \, .
\eea
Then, using Eqs. (\ref{dr}) and (\ref{nu}), one has
\bea
&&dr = \frac{Q(r,\lambda) \Delta_r}{\Xi^4R^2\Sigma}\left[\Delta_\theta dr_* -P^2(\theta,\lambda)\nu d\lambda \right] \, , \label{dr2} \\
&&d\theta = \frac{P(\theta,\lambda)\Delta_\theta}{\Xi^4R^2\Sigma}\left[\Delta_rdr_* +Q^2(r,\lambda)\nu d\lambda \right] \, . \label{dtheta}
\eea
Since the outgoing and ingoing null congruences are defined to be $du = dv = 0$, which means $dr_*^2 = dt^2$, the metric of the Kerr-Newman-AdS$_4$ black hole on the null hypersurface is reduced to
\be\label{dh2}
dh^2 = R^2\sin^2\theta(d\phi -\Omega dt)^2 +\frac{\nu^2P^2(\theta,\lambda)Q^2(r,\lambda)}{\Xi^4R^2}d\lambda^2 \, .
\ee
Because  the volume element of a null hypersurface is the square root of the determinant of the induced metric, the points where the induced metric determinant goes to zero correspond to the NHC. Thus, the condition of the NHC is
\be\label{NHC}
\nu P(\theta,\lambda)Q(r,\lambda)\sin\theta = 0 \, ,
\ee
which is the determinant of the induced metric (\ref{dh2}) of the Kerr-Newman-AdS$_4$ black hole. Let us consider the ingoing null hypersurface case (same for the outgoing null hypersurface case), namely, $\lambda = {\rm const}$ and a decreasing $r$, as an example to study the NHC. We need to analyze each factor in Eq. (\ref{NHC}).  Since Eq. (\ref{nu}) gives that the decreasing of $\theta$ will lead to the decreasing of $r$  for a fixed $\lambda$, we have $P(\theta,\lambda) > 0$. Therefore, $Q(r,\lambda) = 0$ is a sufficient condition for the existence of the NHC.

\end{appendices}

\end{document}